\documentclass[twocolumn, prc]{revtex4}
\usepackage{graphicx}
\usepackage{epstopdf}
\usepackage{amssymb}
\usepackage[raggedright]{titlesec}

\begin{document}

%\title{\bf Calculation of the Structure and Thermodynamic Properties of the Proto-Strange Quark Star}

\title{\bf Proto-Strange Quark Star Structure}

\author{\textbf{{ Gholam Hossein Bordbar$^{1,2}$
\footnote{Corresponding author. E-mail:ghbordbar@shirazu.ac.ir}},
{Fatemeh Sadeghi$^{1}$}, { Fatemeh Kayanikhoo $^{3}$} and Ahmad Poostforush$^{1}$}}
\affiliation{$^1$ Department of Physics,
Shiraz University, Shiraz 71454, Iran\footnote{Permanent address}\\
$^2$ Department of Physics and Astronomy, University of Waterloo,
200 University Avenue West, Waterloo, Ontario,
N2L 3G1,
Canada\\
  $^3$ Department of Physics, University of Birjand, Birjand, Iran
}

%
%%%%%%%%%%%%%%%%%%%%%%%%%%%%%%%%%%%%%%%%%%%%%%%%%%%%%%%%%%%%%%%%%%%
\begin{abstract}
\textbf{\abstractname:} In this paper, {we investigate the newborn strange quark stars with constant entropy}.
{We also use the MIT bag model to calculate the thermodynamic
properties in two cases; the density-dependent bag constant and the fixed bag constant ($B=90$ $MeV$).}
{We show that the equation of state becomes stiffer by using the density-dependent bag constant and by increasing the entropy}.
{Furthermore, we show that the adiabatic index of the system reaches to $\frac{4}{3}$ at high densities.}
Later, we calculate the structure of a strange quark star using the equation of state and the general relativistic
equations of hydrostatic equilibrium, the Tolman-Oppenheimer-Volkoff (TOV) equations.
We show that the gravitational mass of the star decreases by increasing the entropy and
the maximum gravitational mass is larger when we use the density-dependent bag constant {at fixed central energy density}.
{It is shown that the mass-radius relation for this system obeys $M$ $\propto$ $R^{3}$ for different cases of the calculations.}
{Finally, we show that for a given stellar mass considering the fixed bag constant, the maximum gravitational redshift
of a strange quark star occurs at larger values of  entropy.}
\\


\end{abstract}
\maketitle
 %%%%%%%%%%%%%%%%%%%%%%%%%%%%%%%%%%%%%%%%%%%%%%%%%%%%%%%%%%%%%%%%%%%%%%%%%%%%%%%%%%%%%

\section*{\normalsize 1. Introduction}
Compact stars are created at the end of the life of massive stars. Study of compact objects has always been an important subject in physics and astrophysics
{since} strong conditions such as high densities and temperatures in the core of these objects can be investigated.
{Study of compact objects was started with the discovery of  white dwarfs and the investigation of their thermodynamic properties using Fermi-Dirac statistics \cite{rk1, rk2, rk2a}.}
{Strange quark stars (SQS) are examples of compact objects, the possibility of their formation was raised by Ivanenko and Kurdgelaidze in 1965 \cite{rk3, rk4}.}
Witten in 1989, suggested that the base state of hadrons could be strange matter \cite{rk5}.

{When nuclear matter is at high enough density (more than $10^{15}$ $g/cm^{3}$) in the core of the star,
the nucleons overlap and it is expected that nucleons convert into quarks under a phase transition \cite{rk5a, rk5b, rk5c} when the nuclear matter changes to the quark matter.}
So, there are two possible cases for these stars:
{a compact object which completely consists of quarks (a pure SQS) or a neutron star with a core consisting of quark matter (a hybrid star);
Itoh calculated the maximum gravitational mass of the SQS in 1970 \cite{rk6}, he ignored the weak interaction of quarks in his calculations.}
In 1976, after presentation of the MIT bag model, Brecher and Caporaso calculated the maximum gravitational mass of the hybrid star \cite{rk7}.

{A pure SQS which is made of strange quark matter (SQM) that includes three flavors of quarks (up, down and strange), is more stable than nuclear matter, and
its gravitational mass is proportional to the third power of radius ($M$ $\propto$ $R^{3}$). The maximum gravitational mass is almost twice the mass of the sun and
has a radius of about $10$$-$$12$ $km$ \cite{ rk8}. The electric field of an SQS is about $10^{18}$$-$$10^{19}$ $V/cm$,
 and it has a strong magnetic field about $10^{15}$$-$$10^{19}$ $G$ \cite{rk9, rk10, rk11}.
The objects, RX-J185635-3754 ($M=1.7$$ M_{\odot}$ and $R=11$ $km$) and 3C58 are candidates for SQS \cite{rk12}.}

{As we know, in the core of massive stars nuclear processes continue in such a way that the final element is $^{56}Fe$. At this point, the degenerate pressure is not enough to keep the star from gravitational collapse,
and the star's life comes to an end.} During the collapse, it is expected that a thin shell of the star is driven out by transition of neutrinos and electromagnetic waves
while the central compact core makes a compact object. In the stage of collapsing, neutrinos are trapped so the collapse is adiabatic.
In other words, the newborn quark star is opaque in the emission of neutrinos with constant entropy \cite{rk13}. Investigation of the protoneutron star matter has been reported in 2002 \cite{rk13a}.
{Before that Bethe et al. had calculated the equation of state in the gravitational collapse of stars \cite{rk14}.}
 Also, Fischer et.al. have studied core-collapse supenovae explosions triggered by a quark-hadron phase transition \cite{rk14a}. Also, the new-born quark stars have been investigated using NJL model by Sadin and Blaschke \cite{rk14b}.
The temperature evaluation of the new-born compact stars have been studied by Wong and Chu \cite{rk14c}.
In another paper Dexheimer et.al. have calculated the properties and structure of the proto-quark stars using both NJL model and MIT bag model \cite{rk14d}.

{We have published different papers regarding strange quark stars where the structural properties of the SQS at zero and finite temperatures have been computed.}
{These include the structure of an SQS using the MIT bag model with the fixed and the density-dependent bag constant in
the presence and the absence of magnetic fields \cite{rk15,rk16,rk17,rk18,rk19}.} We have also computed the maximum gravitational mass and other structural properties of a neutron star with a quark core at
zero temperature and finite temperatures \cite{rk20, rk21}.
Also, we have studied the effect of dynamical quark mass in the calculation of SQS structure using MIT bag model and NJL model at zero temperature \cite{rk22}.

{In this paper, we investigate the thermodynamic properties and the structure of a proto-strange quark star.}
In section $II$, we calculate the thermodynamic properties of SQM using MIT bag model with the constant entropy for different densities.
Also, we investigate the behavior of quarks at high densities.
In section $III$, we compute the maximum gravitational mass and radius of the proto-SQS using the equation of state (EOS) and the Tolman-Oppenheimer-Volkoff (TOV) equations.
Also we calculate the gravitational redshift of the SQS in the last section.

\section*{\normalsize 2. Calculation of the thermodynamic properties of SQM with constant entropy}
\label{II}
{We calculate the thermodynamic parameters of an adiabatic system such as the chemical potential, total energy, and the equation of state (EOS) with the variable temperature.}

{First, we calculate the temperature and chemical potential when entropy is constant. Then we calculate the total energy of the system to find the EOS.}
Also, we use the EOS to calculate the adiabatic index of the system.

{In this paper, we consider the pure SQS containing up, down and strange quarks. In our calculations, the masses of strange quark is considered to be $150$$ MeV/c^2$, while the mass of up and down quarks are ignored.}

\subsection*{\normalsize 2.1. Temperature and chemical potential}
{We use the density and entropy equations of Fermions to calculate the temperature and the chemical potential.}
 Fermi-Dirac distribution function is,
\begin{equation}\label{01}
f_{i}(p)=\frac{1}{e^{\beta{(\epsilon_{i}(p)-\mu_{i})}}+1},
\end{equation}
{where $\mu_{i}$ is the chemical potential of quark $i$, $\beta$ is equal to $1/(k_{B}T)$ with $k_{B}$ as the Boltzmann constant.}
Also $\epsilon$ is the energy of the single-particle,
\begin{equation}\label{02}
\epsilon=(\hbar^{2}k^{2}c^{2}+m^{2}c^{4})^{1/2}-mc^{2}.
\end{equation}
The number density of quark $i$ is,
\begin{equation}\label{03}
n_{i}=\frac{N_{i}}{V_{i}} = \frac{6}{(2\pi)^{3}} \int_{0}^{\infty} f(n_{i},k_{i},T) d^{3}k_{i},
\end{equation}
where $d^3k_{i}=4\pi k^{2} dk$ and $i=u, d, s$. Also the baryonic density is defined in the following form,
\begin{equation}\label{04}
n_{B}=\frac{1}{3} (n_{u}+n_{d}+n_{s}).
\end{equation}
{The entropy per baryon for each type of quark $i$ is given by},
\begin{eqnarray}\label{05}
s_{i}={} &-\frac{3}{n_{B} \pi^{3}} \int_{0}^{\infty} k^{2} (f(n_{i},k_{i},T)\ln{f(n_{i},k_{i},T)}\\  \nonumber
          &+ ((1-f(n_{i},k_{i},T))\ln{(1-f(n_{i},k_{i},T))})dk,
\end{eqnarray}
and the total entropy per baryon of SQM is,
\begin{equation}\label{06}
S_{tot}= s_{u}+s_{d}+s_{s}.
\end{equation}

{We have shown the variation of temperature versus density for different entropies in Fig. \ref{01} where
it can be seen that the temperature increases by increasing entropy. When entropy is fixed, temperature increases as a function of density.}
\begin{figure}
\centering
\includegraphics[width=10cm, height=7cm]{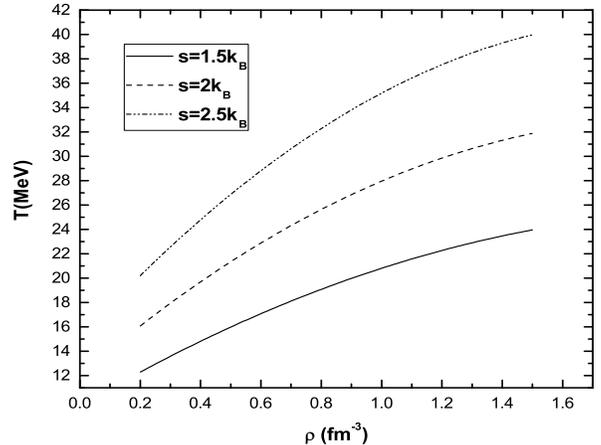}
\caption{Temperature versus baryonic density in different entropies.} \label{01}
\end{figure}
The chemical potential of quarks versus density has been plotted in Fig. \ref{02}.
Since, the chemical potential of down and strange quarks are equal \cite{rk3}, the plot for down quark is not shown.
It can be seen that {for a fixed density, the chemical potential decreases by increasing entropy}.
Also, the chemical potential is a rising function of density whenever entropy is fixed.
\begin{figure}
\centering
\includegraphics[width=10cm, height=7cm]{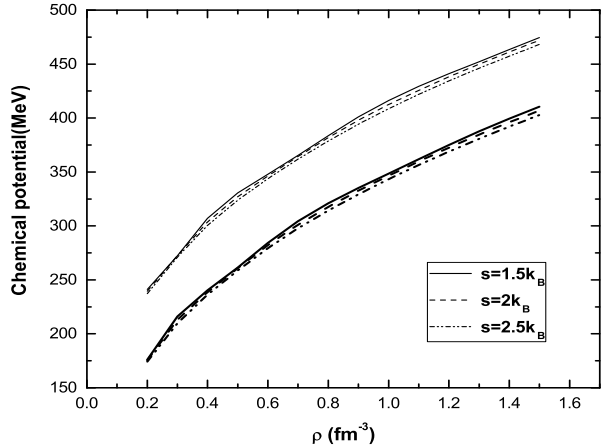}
\caption{The chemical potential of up and strange quarks versus density in different entropies. Thin curves (thick curves) have been used for up quarks (strange quarks).} \label{02}
\end{figure}
\subsection*{\normalsize 2.2. Energy of SQM}
We use the MIT bag model to calculate the total energy of SQM. The MIT bag model is a model describing quarks bounded in the hadrons.
In this model, quarks are considered like a gas in a bag with pressure of $B_{bag}$. The total energy of SQM ($\epsilon_{tot}$) is the kinetic energy of quarks plus the bag constant.
We consider two cases for the bag constant: $A$$)$ fixed bag constant equal to $90$ $MeV/fm^{3}$ and $B$$)$ and density-dependent bag constant which is defined by a Gaussian equation using the experimental data of CERN \cite{rk13, rk16,rk23},
\begin{equation}\label{07}
B_{n}= B_{\infty}+(B_{0}+B_{\infty})e^{-\beta (\frac{n}{n_{0}})^{2}},
\end{equation}
where $\beta$ is the numerical parameter equal to the density of normal nuclear matter ($n_{0}=0.17$), and $B_{0}= B(n=0)$ is equal to $400$ $MeV/fm^{3}$.
Also, the value of the parameter $B_{\infty}$  depends only on the value of the parameter $B_{0}$ and is obtained  by the  LOCV cluster expansion.
We found that $B_{\infty}$  is $8.99$$ MeV/fm^3$ where the energy density of quark matter is equal to that of the nuclear matter \cite{rk19}.

The total energy of SQM using the MIT bag model is defined as follows,
\begin{equation}\label{A8}
\epsilon_{tot}= \varepsilon_{tot}+B_{bag}.
\end{equation}
The kinetic energy density $\varepsilon_{i}$ is,
\begin{equation}\label{09}
\varepsilon_{i}= \frac{3}{\pi^{2}} \int_{0}^{\infty} \epsilon(k_{i}) f(n_{i},k_{i},T) k_{i}^2 dk_{i},
\end{equation}
and the total kinetic energy of SQM is,
\begin{equation}\label{10}
\varepsilon_{tot}= \varepsilon_{u}+\varepsilon_{d}+\varepsilon_{s}.
\end{equation}
In Fig. \ref{03}, the energy density of SQM, by considering fixed bag constant ($B=90 MeV/fm^{3}$) and density-dependent bag constant in different entropies has been plotted.
It can be seen that in both cases ($B= 90 MeV/fm^{3}$ and density-dependent bag constant), the energy density increases by increasing density in the fixed entropy case.
Also, it is shown that in the fixed density case, the energy density increases by increasing entropy. This behavior is due to the increase in temperature and chemical potential of SQM.
By considering $B= 90 MeV/fm^{3}$, the energy density curve is almost linear but by considering the density-dependent bag constant, the curve has a minimum point of energy density at the density of $0.4$$fm^{-3}$.
The minimum in energy density is a sign of a stable state. {Furthermore, it can be seen that when we use the density-dependent bag constant, the energy is larger than the case when we use the fixed bag constant.}

\begin{figure}
\centering
\includegraphics[width=10cm, height=7cm]{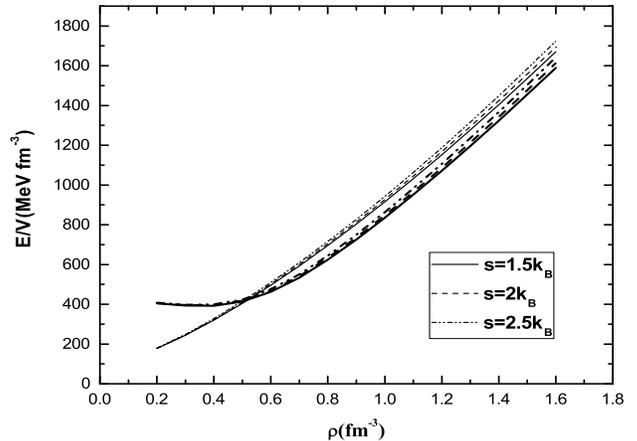}
\caption{The energy density versus density of system in different entropies using $B= 90 MeV/fm^{3}$ (thin curves) and density-dependent bag constant (thick curves).}
\label{03}
\end{figure}

\subsection*{\normalsize 2.3. The equation of state}

{The equation of state is given by the following relation,}
\begin{equation}\label{A11}
P= n\frac{d\epsilon_{tot}}{dn}-\epsilon_{tot},
\end{equation}
where $n$ is the density and $\epsilon_{tot}$ is the total energy density. {We have calculated the EOS of the system using the Eqs. \ref{A8} and \ref{A11}.}

Fig. \ref{04} shows pressure of the system versus density in different entropies using fixed bag constant and density-dependent bag constant.
In both cases, the pressure increases as a function of density. {Also, we see that the pressure increases by increasing the entropy at each value of the given density }for the case of $B= 90 MeV/fm^{3}$, while for case of density-dependent bag constant, {this behavior is seen after a particular value of  density.}
As it is shown in Fig. \ref{04}, in the case of  $B= 90 MeV/fm^{3}$, {for densities greater than $0.8 fm^{-3}$},  the equation of state is softer than the case of density-dependent bag constant.
{These behaviors are due to the change of bag constant versus the density in the case of density dependent bag constant (Eq. \ref{07}).}
Also, it is clear that each pressure curve for $B= 90 MeV/fm^{3}$ is almost a linear function of density while for density-dependent bag constant this is not correct.
{The results for the case of density-dependent bag constant is more similar to that we expected for the fermions.}

\begin{figure}[h]
\begin{center}$
\begin{array}{cc}
\includegraphics[width=4.6cm]{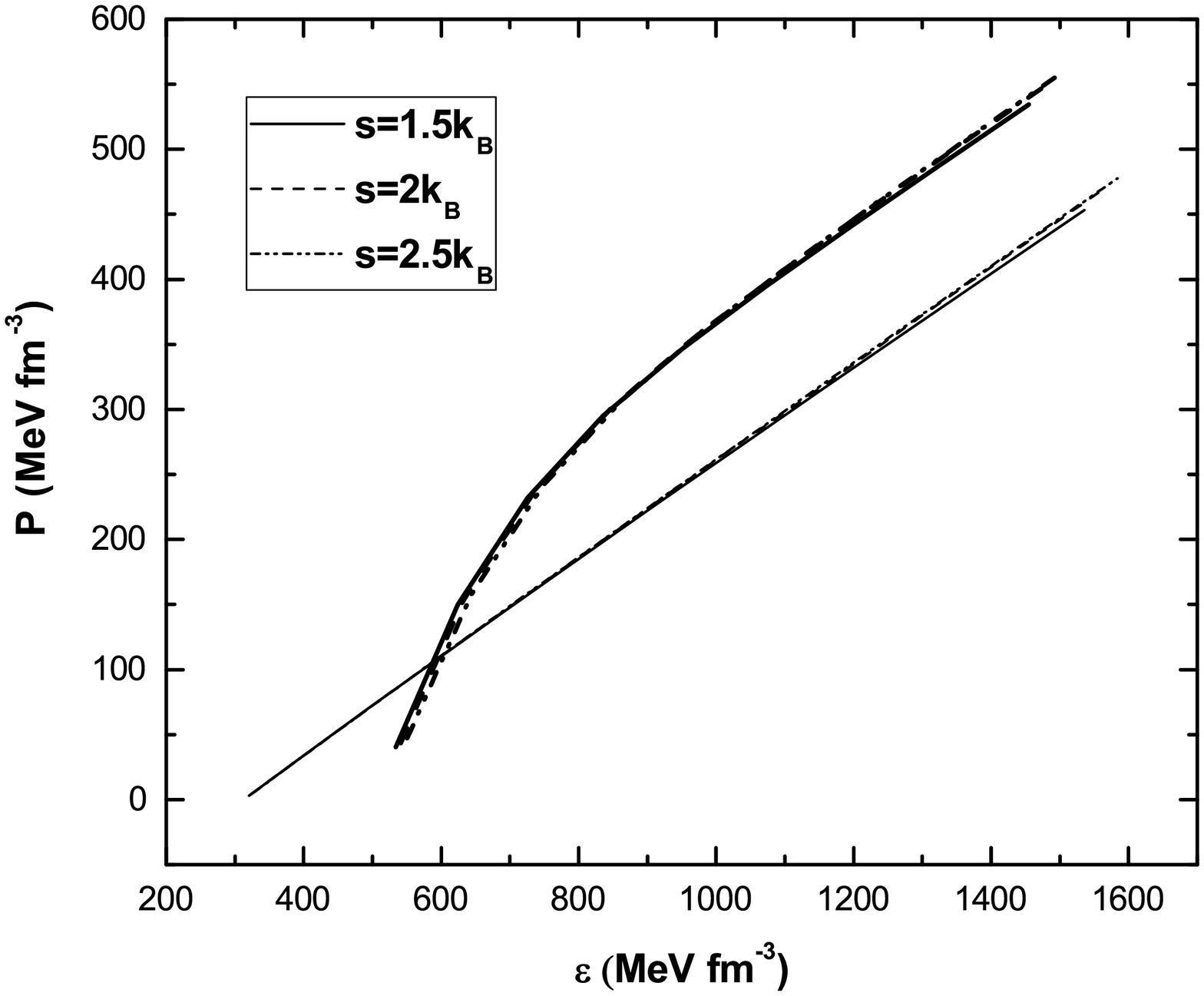}
\includegraphics[width=4.6cm]{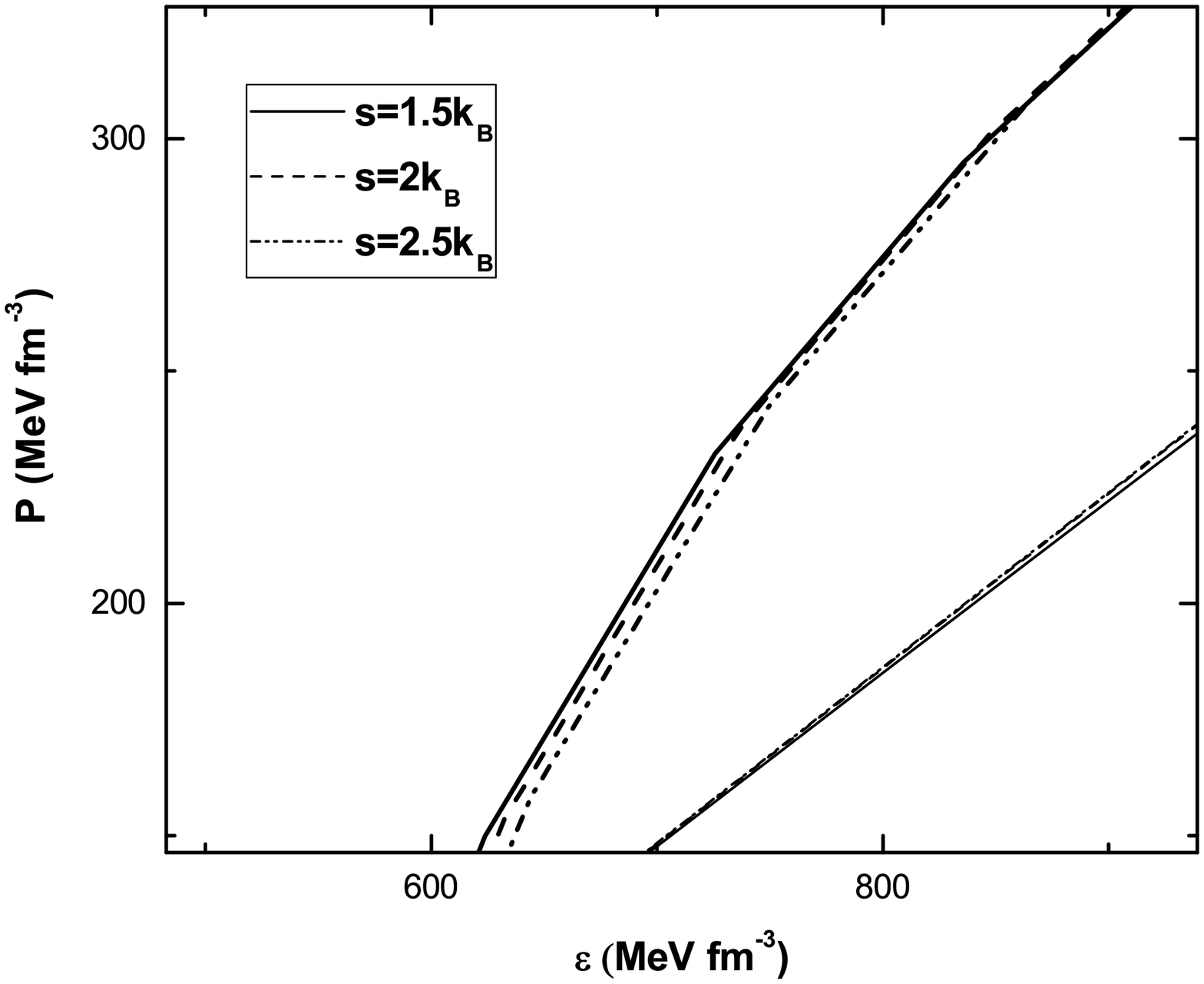}
\end{array}$
\end{center}
\caption{The pressure versus the energy density in different entropies using $B= 90 MeV/fm^{3}$ (thin curves) and density-dependent bag constant (thick curves).} \label{04}
\end{figure}

\subsection*{\normalsize 2.4. Calculation of the adiabatic index using the polytropic behavior}

{The polytropic behavior of a system is one of the important topics in thermodynamics governed by an equation of the form,}
\begin{equation}\label{11}
P \propto n^{\gamma},
\end{equation}

where $P$ is pressure, $n$ is density, and $\gamma$ is the polytropic factor.
{For adiabatic systems, $\gamma$ is the adiabatic constant. The adiabatic constant is the parameter which shows the stiffness of system and is defined as a function of density,}
\begin{equation}\label{12}
\gamma=\frac{dln(P)}{dln(n)}.
\end{equation}

The logarithmic variations of pressure versus density has been plotted in Fig. \ref{05} for both cases, $B= 90 MeV/fm^{3}$ and density-dependent bag constant with different entropies.
It is seen that in both cases, the curves are rising. {Also, at fixed density, the  logarithmic variations of pressure increases by increasing entropy,
furthermore, at lower densities, the gradient of the graph is steeper. The logarithmic variations of pressure using the density-dependent bag constant is larger at densities higher than $0.8 fm^{-3}$.}

\begin{figure}
\centering
\includegraphics[width=9cm, height=7cm]{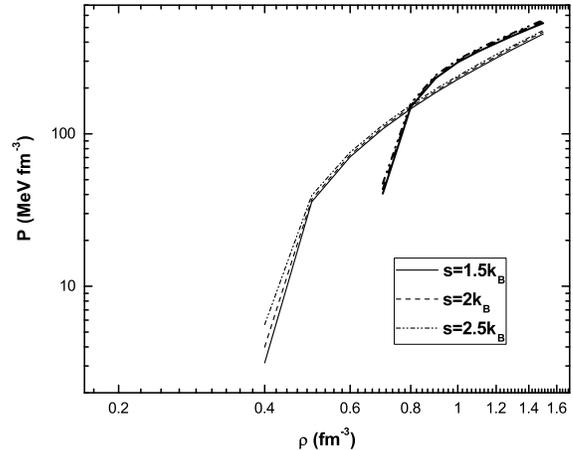}
\caption{The  logarithmic variations of pressure versus density in different entropies using $B= 90 MeV/fm^{3}$ (thin curves) and density-dependent bag constant (thick curves).} \label{05}
\end{figure}

{We have plotted the adiabatic index as a function of density in Fig. \ref{06} with different entropies for both $B= 90 MeV/fm^{3}$ and density-dependent bag constant.
It is clear that the adiabatic index decreases to $\frac{4}{3}$ by increasing density in both cases of bag constants.
Since by increasing the density, interaction of quarks decrease, it is excepted that the system behaves as a free Fermi gas at high densities so the adiabatic index reaches $\frac{4}{3}$.
It is also seen that the adiabatic index decreases by increasing the entropy at a given density, since the equation of state becomes stiffer as the entropy increases. This behavior occurs in both cases considered for bag constant.}

\begin{figure}
\centering
\includegraphics[width=10cm, height=7cm]{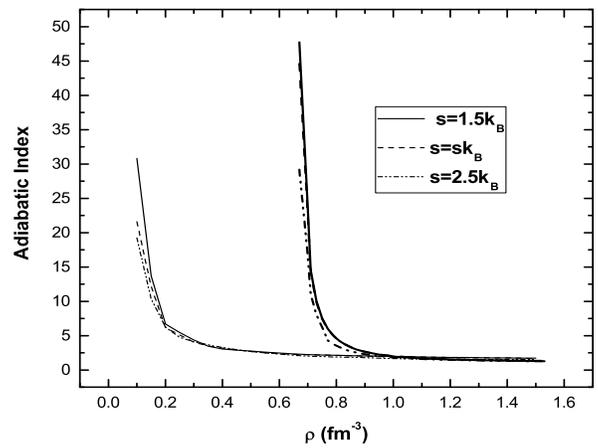}
\caption{The adiabatic index versus density in different entropies using $B= 90 MeV/fm^{3}$ (thin curves) and density-dependent bag constant (thick curves).} \label{06}
\end{figure}

\section*{\normalsize 3. The structure of the proto-strange quark star}
{Quark stars are relativistic objects, so we should use general relativity to calculate their structure.} To investigate the structure of the SQS one needs to calculate the mass and radius of the star.
The mass- radius relation can be obtained using Tolman-Oppenheimer-Volkoff (TOV) equations,
\begin{equation}\label{13}
\frac{dP}{dr}=-\frac{G\left[\varepsilon(r)+\frac{P(r)}{c^{2}}\right]\left[m(r)+\frac{4\pi
r^{3}P(r)}
{c^{2}}\right]}{r^{2}\left[1-\frac{2Gm(r)}{rc^{2}}\right]},
\end{equation}
\begin{equation}\label{14}
\frac{dm}{dr}=4\pi r^{2}\varepsilon(r).
\end{equation}
Using the equation of state obtained in the previous
section and the boundary conditions ($P(r=0)=P_{c}$, $P(r=R)=0$, $m(r=0)=0$ and $m(r=R)=M_{max}$) we integrate the TOV equations to compute the structure of the
SQS \cite{rk24, rk25}. $M_{max}$ is the maximum gravitational mass of the star and the corresponding radius (the radius of the star) is shown as $R$.
The maximum gravitational mass is important because it shows the star's stability against becoming a black hole.

{The changes in gravitational mass} versus the central energy density is shown in Fig. \ref{07} for both cases of bag constant in different entropies.
It can be seen that the gravitational mass increases as a function of the central energy density until the maximum mass is reached.
{It is also clear that in both cases considered for bag constat, the maximum gravitational mass decreases by increasing the entropy {at a fixed central energy density}.
The maximum gravitational mass has a larger value when we have used the density-dependent bag constant.}

\begin{figure}
\centering
\includegraphics[width=10cm, height=7cm]{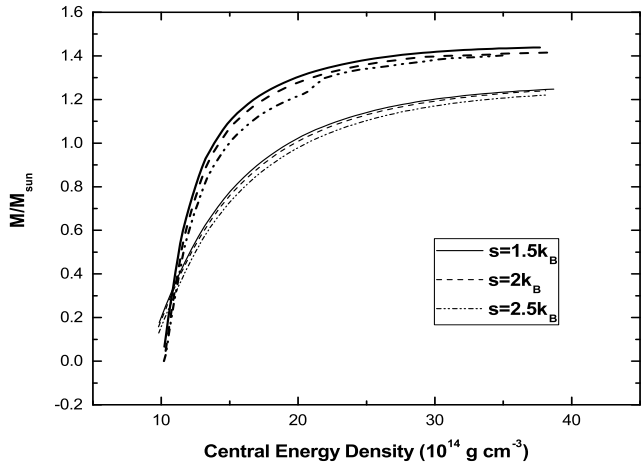}
\caption{The gravitational mass versus central energy density in different entropies using $B= 90 MeV/fm^{3}$ (thin curves) and density-dependent bag constant (thick curves).} \label{07}
\end{figure}

\begin{figure}
\centering
\includegraphics[width=10cm, height=7cm]{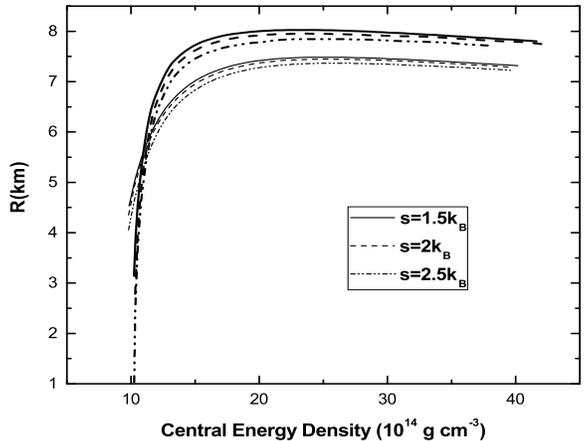}
\caption{The radius versus central energy density in different entropies using $B= 90 MeV/fm^{3}$ (thin curves) and density-dependent bag constant (thick curves).} \label{08}
\end{figure}

In Fig. \ref{08}, the radius of the SQS versus the central energy density has been plotted for both cases of bag constant in different entropies.
It is shown that the radius increases by increasing the central energy density until it reaches the maximum value of the radius. It is also shown that for each given central energy density, the maximum value of radius decreases by increasing the entropy.
{Furthermore, the radius of the SQS has a larger value when we consider the density-dependent bag constant.}

\begin{figure}
\centering
\includegraphics[width=10cm, height=7cm]{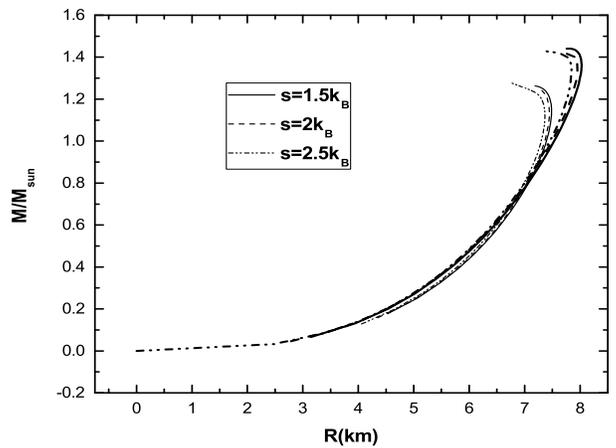}
\caption{The gravitational mass versus radius in different entropies using $B= 90 MeV/fm^{3}$ (thin curves) and density-dependent bag constant (thick curves).} \label{09}
\end{figure}
 In Fig. \ref{09}, {the gravitational mass is also shown as a function of the radius with the same conditions as of the previous figures.}
It can be seen that the larger gravitational mass for any fixed radius is related to the larger entropy for each case of the bag constant,
and values of mass and radius are larger where we have used the density-dependent bag constant.
Also, in the case of fixed entropy, the gravitational mass increases by increasing the radius, but after reaching the maximum gravitational mass the radius decreases.
In fact according to our calculations, {the star is more stable for the case of density-dependent bag constant.}
Increasing radius of SQS as a function of the gravitational mass is different from that of a neutron star.
{For the SQS, our structure results show that the relation of mass-radius follows $M$ $\propto$ $R^{3}$ in each considered case.}

\subsection*{\normalsize 3.1. Gravitational redshift}

{The gravitational redshift $Z$ of stars is defined by the following relation,}
\begin{equation}\label{15}
Z=(1-\frac{2M}{Rc^{2}})^{-1/2}-1.
\end{equation}
{The gravitational redshift of SQS has been plotted in Fig. \ref{10} with the same conditions as those of the previous figures.} It is clear that the gravitational redshift increases by increasing entropy.
Also, the gravitational redshift has a larger value by considering the density-dependent bag constant.
{The maximum value of the gravitational redshift for SQS, $Z=0.48$, is related to a system with the density-dependent bag constant with the entropy $S=2.5k_B$, and the minimum value, $Z=0.4$,
is related to the fixed bag constant with the entropy $S=1.5k_B$.}
\begin{figure}
\centering
\includegraphics[width=10cm, height=7cm]{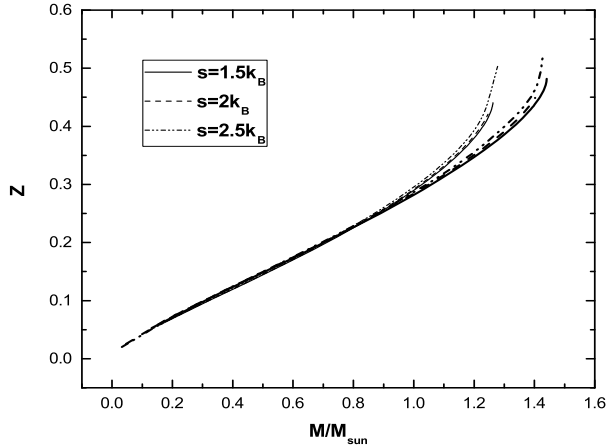}
\caption{The gravitational redshift versus the gravitational mass in different entropies using $B= 90 MeV/fm^{3}$ (thin curves) and density-dependent bag constant (thick curves).} \label{10}
\end{figure}
%

%%%%%%%%%%%%%%%%%%%%%%%%%%%%%%%%%%%%%%%%%%%%%%%%%%%%%%%%%%%%%%%%%%%%%%%%%%%%%
%%%%%%%%%%%%%%%%%%%%%%%%%%%%%%%%%%%%%%%%%%%%%%%%%%%%%%%%%%%%%%%%%%%%%%%%%%%%%

\section*{\normalsize 4. Conclusions}

{In this paper, we have investigated thermodynamic properties and structure of proto-strange quark star (SQS)
in which the neutrinos are trapped, and therefore the entropy is constant.
In section $II$, we calculated the temperature and chemical potential of strange quark matter (SQM) with fixed entropy.
Next, we calculated the free energy of  SQM in different entropy cases. We have used the MIT bag model with two cases of bag constants
(fixed bag constant $B=90$ $MeV/fm^3$ and density-dependent bag constant).
Using the free energy we have calculated the equation of state (EOS) of the SQM.
We showed that the EOS of the system becomes stiffer by considering the density-dependent bag constant. Also we showd that the EOS is softer in the lower entropies.
Furthermore, we investigated the adiabatic index of the SQM where we have shown that the adiabatic index has reached $\frac{4}{3}$ at high densities
for both cases of the bag constants. Actually the SQM behaves like a free fermi gas at higher densities because {the bag constant decreases in comparison with
the energy scale}.
In section $III$, we used the EOS and TOV equations to calculate the mass and radius of the SQS.
we showed that the maximum gravitational mass and the radius of SQS decrease by increasing entropy.
In general, we have found that the larger values of entropy lead to the stiffer equation of state, the smaller adiabatic index, and the smaller values of the maximum gravitational mass and radius for the SQS {at a
given central energy density}.
Also, it has been shown that the maximum gravitational mass and radius of the SQS are larger when we use the density-dependent bag constant.
Furthermore, we plotted the gravitational redshift versus the gravitational mass where we have shown that the maximum gravitational redshift occurs
for $S=2.5k_B$ considering the fixed bag constant.
Finally, an $M$ $\propto$ $R^{3}$ relation was found for the mass and radius of SQS for all relevant calculations.}

%%%%%%%%%%%%%%%%%%%%%%%%%%%%%%%%%%%%%%%%%%%%%%%%%%%%%%%%%%%%%%%%%%%%%%%%%%%%%%%%%%%%%%%%%%%%%%%%%%
\begin{acknowledgements}
{We wish to thank Shiraz University Research Council.}
\end{acknowledgements}
%%%%%%%%%%%%%%%%%%%%%%%%%%%%%%%%%%%%%%%%%%%%%%%%%%%%%%%%%%%%%%%%%%%%%%%%%%%%%%%%%%%%%%%%%%%%%%%%%%

%%%%%%%%%%%%%%%%%%%%%%%%%%%%%%%%%%%%%%%%%%%%%%%%%%%%%%%%%%%%%%%%%%%%%%%%%%%%%%%%%%%%%%%%%%%%%%%%%%%%%%

%%%%%%%%%%%%%%%%%%%%%%%%%%%%%%%%%%%%%%%%%%%%%%%%%%%%%%%%%%%%%%%%%%%%%%%%%%%%%%%%%%%%%%%%

%%%%%%%%%%%%%%%%%%%%%%%%%%%%%%%%%%%%%%%%%%%%%%%%%%%%%%%%%%%%%%%%%%%%%%%%%%%%%%%%%%%%%%%%%%%

\begin{thebibliography}{34}
\section*{\normalsize References}
\bibitem{rk1}  E Ostgaard \emph{Phys. Rep.} {\bf 242} 4 (1994)
\bibitem{rk2} M Camenzind \emph{Compact Objects in Astrophysics} (Springer Berlin Heidelberg) (2007)
\bibitem{rk2a} R K Pathria \emph{Statistical Mechanics} (Pergamon Press) (1980)
\bibitem{rk3} D D Ivanenko and D F Kurdgelaidze \emph{Astrophys.} {\bf 1} 251 (1965)
\bibitem{rk4} D D Ivanenko and D F Kurdgelaidze \emph{Lett. Nuov. Cim.} {\bf 2} 13 (1969)
\bibitem{rk5} E Witten \emph{Phys. Rev.} {\bf D30} 272 (1984)
\bibitem{rk5a} K Nakazato, K Sumiyoshi and S Yamada \emph{Astron. Astrophys.} {\bf A50} 558 (2013)
\bibitem{rk5b} K Nakazato, K Sumiyoshi and S Yamada \emph{Phys. Rev.} {\bf D77} 103006 (2008)
\bibitem{rk5c} K Nakazato, K Sumiyoshi and S Yamada \emph{Astrophys. J.} {\bf 721} 1284 (2010)
\bibitem{rk6} N Itoh \emph{Prog. Theor. Phys.} {\bf 44} 291 (1970)
\bibitem{rk7}  K Brecher and G Caporaso \emph{Nature} {\bf 259} 377 (1976)
\bibitem{rk8}  F Ozel \emph{ Nature} {\bf 441} 1115 (2006)
\bibitem{rk9} F Weber, M Orsaria, H Rodrigues and S H Yang \emph{Proceedings of the International Astronomical Union} {\bf 8} 61 (2012)
\bibitem{rk10} M Bocquet, S Bonazzola, E Gourgoulhon and  J Novak \emph{Astron. Astrophys.} {\bf 301} 757 (1995)
\bibitem{rk11}  M Malheiro, S Ray, H J Mosquera Cuesta and J Dey \emph{Int. J. Mod. Phys.} {\bf D16} 489499 (2007)
\bibitem{rk12} M Prakash et al. \emph{Nucl. Phys.} {\bf A715}, 835c (2003)
\bibitem{rk13} S Shapiro and S Teukolsky  \emph{Black Holes, White Dwarfs and Neutron Stars} (Wiley, New York) (1983)
\bibitem{rk13a} G H Bordbar \emph{Int. J. Theor. Phys.} {\bf 41} 309 (2002)
\bibitem{rk14} H A Bethe et al. \emph{Nucl. Phys.} {\bf A324} 487 (1979)
\bibitem{rk14a}  T Fischeret al. \emph{Astrophys. J.} {\bf 194} 39 (2011)
\bibitem{rk14b} F Sandin and D Blaschke \emph{Phys. Rev.} {\bf D75} 125013 (2007)
\bibitem{rk14c} K W Wong  and M C Chu \emph{Month. Not. Roy. Astron. Soc.} {\bf 350} 42 (2004)
\bibitem{rk14d} V Dexheimer, J R Torres,  D P Menezes \emph{Eur. Phys. J.} {\bf C73} 2569 (2013)
\bibitem{rk15} G H Bordbar  and A Peivand \emph{Res. Astron. Astrophys.} {\bf 11} 851 (2011)
\bibitem{rk16} G H Bordbar, A Poostforush and A Zamani \emph{Astrophys.} {\bf 54} 277 (2011)
\bibitem{rk17} G H Bordbar, H Bahri and  F Kayanikhoo \emph{Res. Astron. Astrophys.} {\bf 12} 1280 (2012)
\bibitem{rk18} G H Bordbar,  F Kayanikhoo and  H Bahri \emph{Iranian J. Sci. Tech.} {\bf A37} 165 (2013)
\bibitem{rk19} G H Bordbar and Z Alizadeh \emph{Astrophys.} {\bf 57} 130 (2014)
\bibitem{rk20} G H Bordbar, M Bigdeli and T Yazdizadeh \emph{Int. J. Mod. Phys.} {\bf A21} 5991 (2006)
\bibitem{rk21} T Yazdizadeh and G H Bordbar \emph{Res. Astron. Astrophys.} {\bf 11} 471 (2011)
\bibitem{rk22} G H Bordbar and B Ziaei \emph{Res. Astron. Astrophys.} {\bf 12} 540 (2012)
\bibitem{rk23} H Li, X L Luo and H S Zong \emph{Phys. Rev.} {\bf D82} 065017 (2010)
\bibitem{rk24}  R Kjelsberg \emph{The Cooling of Neutron Stars} (Lulu publication) (2012)
\bibitem{rk25} P Haensel, A Y Potekhin and D G Yakovlev \emph{Neutron stars 1: Equation of state and structure} (Springer) (2007)


%
\end{thebibliography}
\end{document}